\documentclass{ieeeaccess}
\usepackage{cite}
\usepackage{amsmath,amssymb,amsfonts}
\usepackage{algorithmic}
\usepackage{graphicx}
\usepackage{tabularx}
\usepackage{booktabs}
\usepackage{caption}
\usepackage{textcomp}
\usepackage{soul}  
\usepackage{xcolor}  
\usepackage{soul}  
\usepackage{tabularx}

\sethlcolor{yellow}  
\def\BibTeX{{\rm B\kern-.05em{\sc i\kern-.025em b}\kern-.08em
    T\kern-.1667em\lower.7ex\hbox{E}\kern-.125emX}}
\begin{document}
\history{Date of publication xxxx 00, 0000, date of current version xxxx 00, 0000.}
\doi{10.1109/ACCESS.2017.DOI}

\title{Compound Mask for Divergent Wave Imaging in Medical Ultrasound}
\author{\uppercase{Zahraa Alzein}\authorrefmark{1},
\uppercase{Marco Crocco\authorrefmark{2}, and Daniele D. Caviglia
}\authorrefmark{1}
}
\address[1]{Dipartimento di Ingegneria Navale, Elettrica, Elettronica e delle Telecomunicazioni (DITEN), Università degli Studi di Genova, Italy (e-mail: zahraa.alzein@edu.unige.it) }
\address[2]{Esaote S.p.A, Genoa, Italy (e-mail: marco.crocco@esaote.com)}

\markboth
{Author \headeretal: Preparation of Papers for IEEE TRANSACTIONS and JOURNALS}
{Author \headeretal: Preparation of Papers for IEEE TRANSACTIONS and JOURNALS}

\corresp{Corresponding author: Zahraa Alzein (e-mail: zahraa.alzein@edu.unige.it).}

\begin{abstract}

Divergent wave imaging with coherent compounding allows obtaining broad field-of-view and higher frame rate with respect to line-by-line insonification. However, the spatial and contrast resolution crucially depends on the weights applied in the compound phase, whose optimization is often cumbersome and based on trial and error. This study addresses these limitations by introducing a closed-form approach that maps the transmit apodization weights used in synthetic aperture imaging into the compound mask applied to divergent wave imaging. The approach draws inspiration from a successful technique developed for plane wave imaging, leveraging synthetic aperture imaging as a reference due to its superior image quality. It works for both linear and convex geometries and arbitrary spatial arrangements of virtual sources generating divergent waves. The approach has been validated through simulated data using both linear and convex probes, demonstrating that the Full Width at Half Maximum (FWHM) in Divergent Wave Linear Array (DWLA) increased by 7.5\% at 20 mm and 9\% at 30 mm compared to Synthetic Aperture Linear Array (SALA). For Divergent Wave Convex Array (DWCA), the increase was 1.64\% at 20 mm and 26.56\% at 30 mm compared to  Synthetic Aperture Convex Array (SACA), witnessing the method’s effectiveness.
\end{abstract}


\titlepgskip=-15pt

\maketitle

\section{Introduction}
\label{sec:introduction}
\IEEEPARstart{D}{ivergent} wave imaging (DWI) is an advanced ultrafast ultrasound technique that enables broad field of view and a higher frame rate with respect to traditional focused transmission and line-by-line reconstruction [1][2]. However, the high frame rate is accompanied by low quality in terms of resolution and contrast because the beamforming process is applied only in the receive mode. This problem can be mitigated by using coherent recombination of echoes related to several divergent-wave transmissions, to recover high-quality images without degrading the high-frame-rate capabilities, as previously proposed for plane wave imaging. This method, originally detailed by Montaldo et al. [3] for plane waves, leverages the coherent summation of echoes from various plane wave transmission angles. The compound method can be extended to divergent waves, by substituting the transmission angles with the positions of the virtual source from which the divergent waves are assumed to originate. A crucial step in coherent compounding is the optimization of the compounded weights, or compound mask, i.e. the weights multiplied by each component before the coherent sum.  

Several studies have focused on applying weights for coherent compounding in plane wave imaging where prior studies have drawn parallels between coherent plane-wave compounding (CPWC) and synthetic transmit aperture imaging (STAI), suggesting that STAI apodization techniques can be adapted for angular weights in CPWC [4]. Other approaches aim to dynamically estimate the compound weights from the received data, for example adopting  Capon's minimum variance beamforming [5]. Recent advancements have incorporated Independent Component Analysis (ICA) into CPWC for angular apodization. ICA decomposes the received signals into statistically independent components, which helps estimate the angular weights more effectively [6]. This method assumes that each plane-wave transmission provides a nonindependent observation of the target field, thus enabling the reconstruction of high-quality images through optimal weight estimation.

However, these adaptive approaches have not been yet investigated for DWI .Various researchers have proposed different compounding strategies for DWI [7]. Two predominant transmit sequences for coherent compounding have been described thus far. One method involves generating a divergent wave using the full aperture, whereby the point spread function (PSF) is rotated between transmits by moving the virtual focus point along an arc at a constant distance from the probe's center [8,9]. The other method uses only part of the aperture to generate the divergent wave, tilting the PSF by sliding the active aperture across the array [10]. It remains uncertain which of these methods yields superior results. According to [7], spatial compounding by shifting the active aperture of a diverging wave source can enhance image quality more effectively than rotating a diverging wave generated with the full aperture. However, this study does not address the optimal weights to apply during coherent compounding. 

From this perspective, this study derives a relationship between an arbitrary apodization law in STAI and the compound mask weights in DWI, thus leveraging established optimization criteria in STAI transmit apodization. In particular, under suitable hypotheses, a closed-form solution is computed allowing straightforward mapping of STAI apodization coefficients into DWI compound mask coefficients. The method works for both linear and convex probe geometries and arbitrary arrangements of virtual source positions, thus allowing the approach to be widely applicable. In silico tests confirm the good agreement between STAI and DWI images once the mapping method is applied.

\noindent The contributions of this study can be summarized as follows:
\begin{itemize}
    \item Derivation of a closed-form approach to compute compound mask weights for divergent wave imaging (DWI), enabling direct mapping from transmit apodization weights used in synthetic aperture imaging (SAI). This method eliminates the need for complex and computationally intensive optimization algorithms.
    \item The proposed approach significantly enhances the spatial and contrast resolution in DWI by optimizing the compound mask weights applied during coherent compounding. This optimization reduces the dependence on trial-and-error methods, offering a systematic solution.
    \item The method is adaptable for both linear and convex probe geometries, supporting various spatial arrangements of virtual sources. This broadens the scope of application, making the method suitable for a range of high-frame-rate ultrasound imaging scenarios.
    \item The compound mask weights can be computed offline, minimizing the computational load during real-time imaging. This reduces memory requirements and computational overhead compared to traditional apodization techniques.
    \item  Validation is performed using simulated data for both linear and convex geometries. Results demonstrate the approach's effectiveness in preserving lateral resolution, closely matching the performance of synthetic aperture imaging (SAI) apodization.
\end{itemize}

The paper is organized as follows. In Section II, the theoretical derivation of the compound mask function for divergent wave imaging (DWI) is presented, along with its relationship to synthetic aperture imaging (SAI) transmit apodization. Section III presents the results and discussion, which includes the simulation setup, parameters used for validation, and a detailed analysis of the results obtained. Section IV discusses the limitations of the study and outlines potential avenues for further optimization and experimental validation. Finally, Section V concludes with a summary of the key findings and contributions of the study, as well as future research directions.

\section{Materials and Methods }
\subsection{Synthetic Aperture Imaging (SA) with Linear and Convex Probes}

Synthetic Aperture Imaging (SA) is used in ultrasound imaging to combine signals from multiple transmit-receive events to form a high-resolution image. SA allows for dynamic focusing in transmission and reception and improves image quality and resolution. The signal reconstructed at any point (x,z) in the imaging field can be mathematically represented by the following equation:
\begin{equation}
S_{SA}(x,z)=\sum_{r=1}^{M}u(x_{r},z_{r})\sum_{t=1}^{N}v(x_{t},z_{t})h_{tr}(\frac{D_{t}}{c_{0}}+\frac{D_{r}}{c_{0}})
\end{equation}
where $u(x_{r},z_{r})$ and $v(x_{t},z_{t})$ are the receive and transmit apodization windows function of probe elements coordinates $x_{r},z_{r}$ and $x_{t},z_{t}$ \footnote{For the sake of simplicity the dependence of apodization windows from the image point coordinates $(x,z)$ has been dropped}, respectively. The transmit apodization $v(x_{t},z_{t})$ represents the relative amplitude transmitted by the element of index $t$,  whose coordinates are given by the couple $x_{t},z_{t}$, while the receive apodization $u(x_{r},z_{r})$ denotes the values that must be multiplied with the signal received by the element of index $r$ whose coordinate is given by the couple $x_{r},z_{r}.$ The summation indices $M$ and $N$ represent the number of active transmit and receive elements, respectively. $h_{tr}$ is the impulse response from transmit element $t$ to receive element $r$, $D_t$ and $D_r$ are the distance between the point $(x,z)$ and the transmit and receive probe elements of index $t$ and $r$ respectively and $c_{0}$ is the medium's sound speed.The representations for both synthetic aperture with linear array (SALA) and convex array (SACA) are shown in Figure 1.
\begin{figure*}[ht]
    \centering
    \includegraphics[width=1\linewidth]{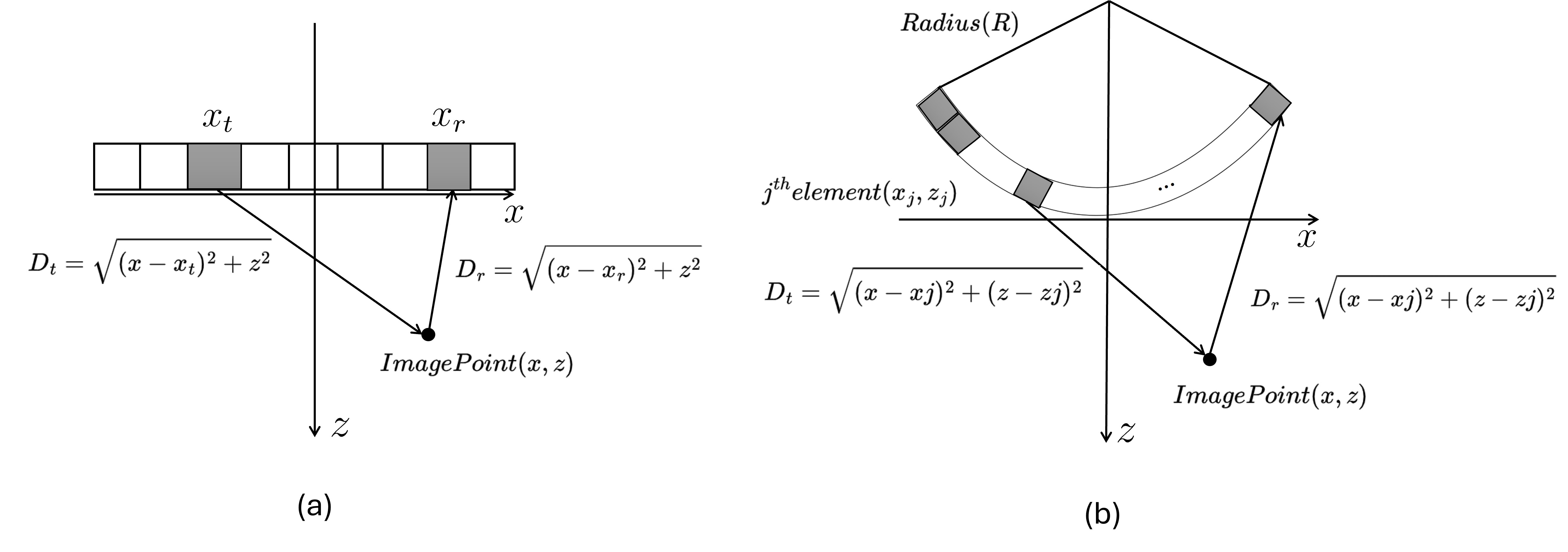}
    \caption{Comparison of SA Linear and Convex Probe Configurations with distances between image point $(x,z)$ and probe elements coordinates in transmission ($x_t,z_t$) and reception ($x_r,z_r$)}.
    \label{fig:enter-label}
\end{figure*}
\subsection{ Divergent Wave Imaging }

A standard approach to generate divergent waves is to consider a virtual source point of coordinates $x_v,z_v$ and compute the relative times of arrival of a spherical wave propagating from the virtual source to each active probe element location. These relative times of arrival are then applied as a transmission delay curve to the probe elements to generate a divergent field approximating the spherical one in the image plane. The effect of angular compounding achieved in plane wave transmission by steering the plane wave angle is achieved here by moving the virtual source in different locations. The optimal placement of virtual sources is a subject of ongoing research but it is outside of the scope of this paper. Given a set of virtual source positions, the signal reconstructed at every point $(x,z)$ can be expressed as follows: 

\begin{equation}
 S_{DW}(x,z)=\sum_{r=1}^{M}u(x_{r},z_r)\sum_{i=1}^{V}w(x_{i},z_{i})\hat{h}_{ir}(\frac{D_{i}}{c_{0}}+\frac{D_{r}}{c_{0}})   
\end{equation}
where $w(x_{i},z_{i})$  denotes the compound mask function of virtual source position and image point coordinates.The compound mask function $w(x_{i},z_{i})$ is defined as the multiplicative weights applied to the beamformed data at each image point $(x,z)$ corresponding to a specific virtual source in transmission [11]. \footnote{For the sake of simplicity, as for apodization, the dependence of compound mask from the image point coordinates $(x, z)$ has been dropped}$\hat{h}_{ir}$ is the two-way impulse response of the divergent wave from the virtual source point $i$ to the image point and back to the receive probe element $r$, and $D_i$ is the distance between the virtual source and the image point.
The impulse response of the divergent wave field can be described as a linear combination of impulse responses related to  each active probe element in transmission with proper relative delays: 
\begin{equation}
    \hat{h}(t)_{ir} = \sum_{t=1}^{N}h_{tr}(t-D_{it}/{c_0})
\end{equation}
where $D_{it}$ is the distance between the virtual source $i$ and the transmit element $t$. 
Substituting the above expression into the equation for the diverging wave imaging one obtains:
\begin{equation}
 S_{DW}(x,z)=\sum_{r=1}^{M}u(x_{r},z_r)\sum_{i=1}^{V}\sum_{t=1}^{N}w(x_{i},z_{i})h_{tr}(\frac{D_{t}}{c_{0}}+\frac{D_{r}}{c_{0}}+\tau_{it})   
\end{equation}
where $\tau_{it}$ is given by:
\begin{equation}
   \tau_{it}=(D_i-D_t-D_{it})/c_0 
\end{equation}

 In Figure 2 the respective distances Di, Dr ,Dit are displayed with
 the linear and convex probe geometries.From the above expression, it can be noticed that both $S_{SA}(x,z)$ and  $S_{DW}(x,z)$  can be thought as a linear combination of two-way probe elements impulse responses $h_{tr}(t)$. This allows, under opportune hypotheses, to derive a simple relationship between the compound mask of  $S_{DW}$ and apodization weights of $S_{DW}$, as described in the next Section.

\begin{figure*}[ht]
    \centering
    \includegraphics[width=1\linewidth]{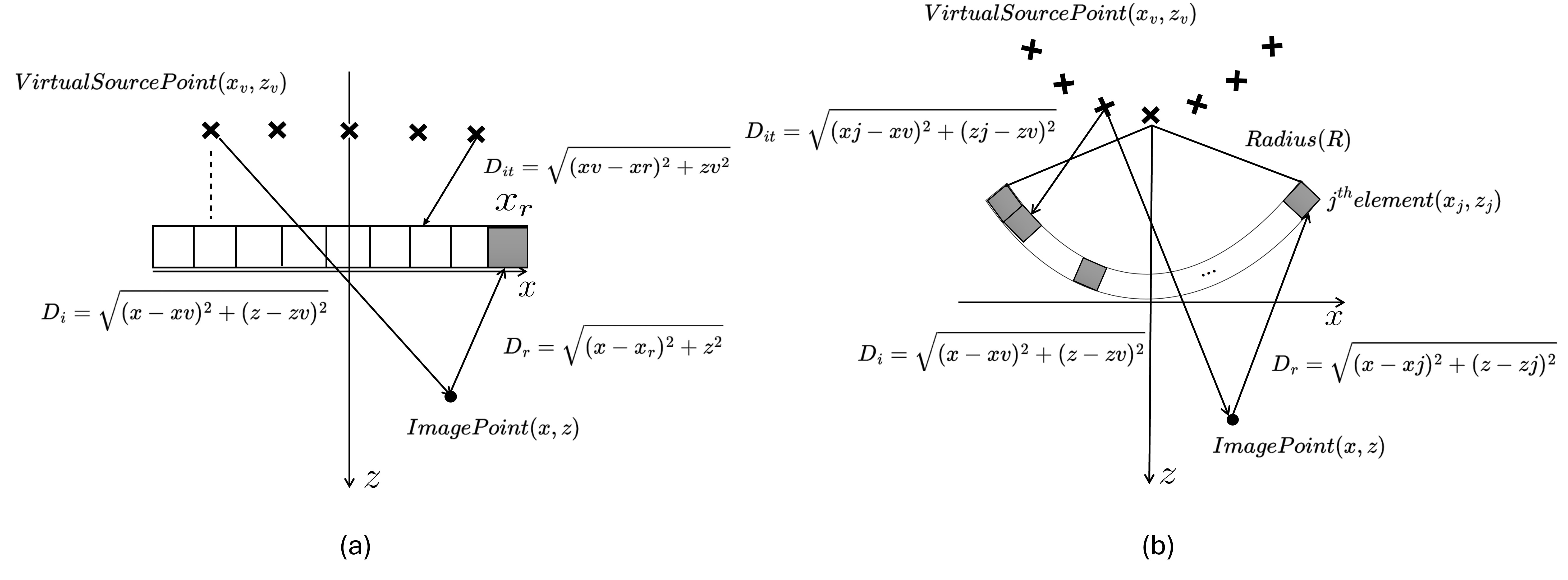}
    \caption{Comparison of DW Linear and Convex Probe Configurations with distances between virtual source $(x_v,z_v)$ and image point $(x,z)$ and probe elements coordinates in transmission ($x_t,z_t$); moreover also the distance between image point and probe elements coordinates in reception ($x_r,z_r$) is displayed.}
    \label{fig:enter-label}
\end{figure*}
\subsection{Relation between transmit apodization in STAI and Compound Mask in DWI}
\begin{figure}
    \centering
    \includegraphics[width=1\linewidth]{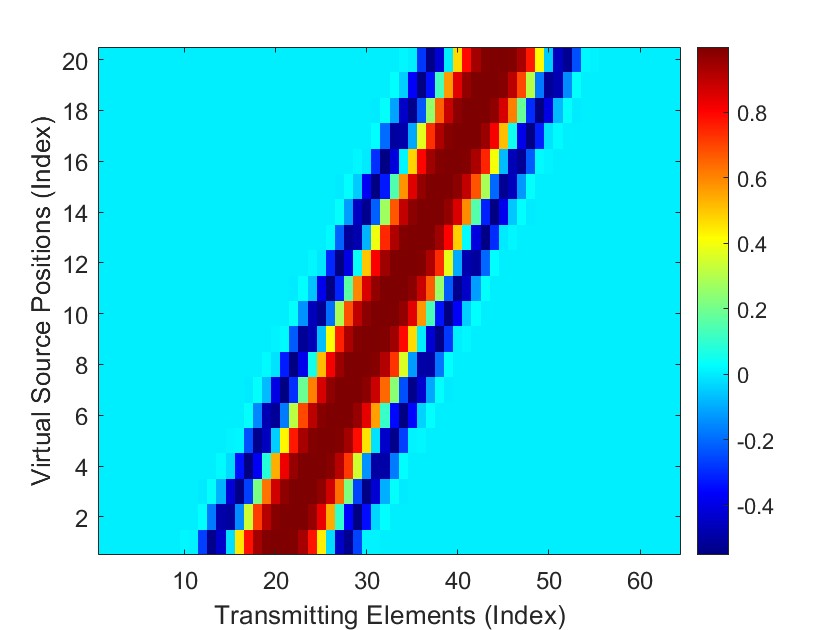}
    \caption{ Examples of values of H matrix in Eq. 7. Row and column indexes are related to virtual sources and probe elements indexes respectively}
    \label{fig:enter-label}
\end{figure}

\begin{figure*}[ht!]
    \centering
    \includegraphics[width=1\linewidth]{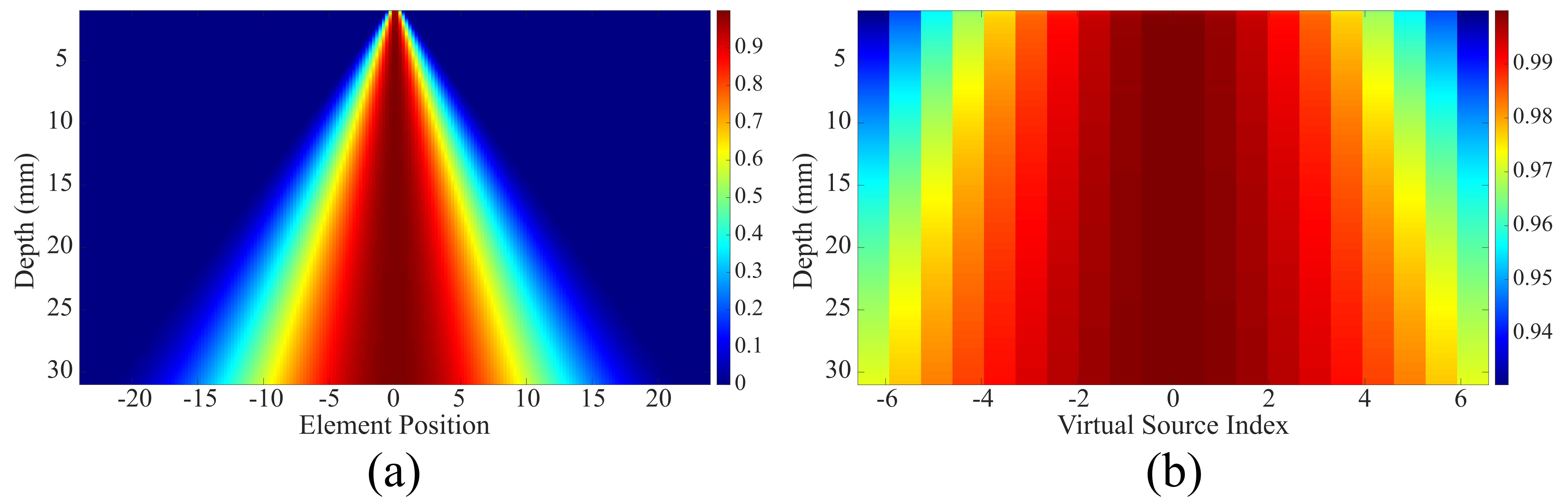}
    \caption{Comparison of Transmit Apodization Profile (a) and Compound Mask Profile (b)}
    \label{fig:enter-label}
\end{figure*}

\begin{figure*}[ht!]
    \centering
    \includegraphics[width=1\linewidth]{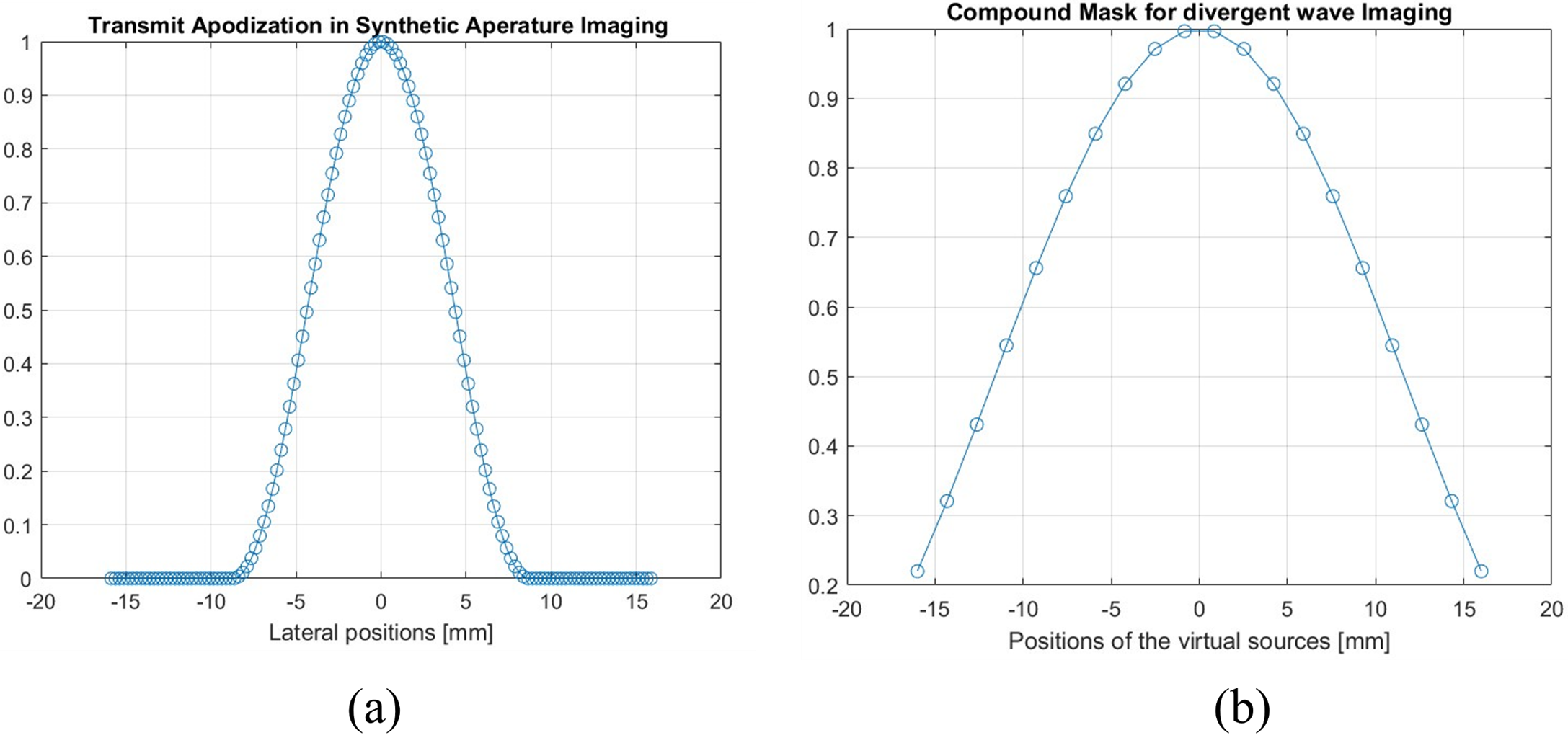}
    \caption{ (a) The transmit apodization $v(x_t)$ in STAI for Hanning window with (b) The corresponding compound mask $w(x_i,z_i)$ in DWLA for a fixed $z_i$}.
    \label{fig:enter-label}
\end{figure*}

In the work of Rodriguez-Molares et al. [4], the authors derived a relationship between apodization in STAI and CPWC, showing that under certain conditions, the angular compounding weights can be derived in closed form from the transmit apodization window of STAI. Here we adapt their approach to divergent waves generated by virtual sources and arbitrary probe geometries. First of all, under the same assumptions of [4] a linear relationship between the compound mask for divergent waves and transmit apodization can be derived assuming that the last expression of $S_{DW}(x,z)$ and the expression for $S_{SA}(x,z)$ are equivalent. The following expression yields: 
\begin{equation} \sum_{i=1}^{V}w(x_i,z_i)H(i,t)=v(x_t,z_t)
\end{equation}
where 
\begin{equation} 
H(i,t)=\frac{E(\tau_{it})}{E(0)}
\end{equation}

where $E(t)$ is the convolution of the transmitted pulse with the two-way electroacoustic transfer function of the probe elements (see Eq. 14 in [4]), collecting the above equations for each apodization weight $v(x_t,z_t)$ a linear system is derived from which the compound mask can be computed, as described in [4].
If we display the matrix $H(i,t)$ in Fig. 3 we can see that only a few elements, corresponding to values of $\tau_{it}$ close to zero, are different from zero. This means that under the additional hypothesis of short-time transmitted pulse, already introduced in [4], it is possible to approximate the matrix $H(i,t)$ in Eq. 6 as diagonal with non-zero values equal to 1. As a consequence, the only non-zero contribution of the sum of Eq. 6 is the one for which $\tau_{it}=0$. In other words equation 6 becomes $w_{x_i,z_i}=v_{(x_t,z_t)}$ for the couple of indexes $(i,t)$ that fulfil the condition $\tau_{it}=0$. 
Looking at the definition of $\tau_{it}$ we see that it represents the difference between the distance from the virtual source to the image point and the sum of the distances between the virtual source and the probe element and between the probe element and the image point, all converted in time. By the triangular inequality, the difference can be zero only if the probe element is aligned along the line connecting the virtual source and the image point. Therefore, we computed the intersection point between the above line and the probe surface, for each virtual source position, thus obtaining the associated probe element and the corresponding apodization weight. For a convex probe, the equation yielding the intersection points ($x_e$,$z_e$) is given by:
\begin{equation}
z_e = \frac{b}{m^2} + \frac{\sqrt{\left( \frac{b}{m^2} \right)^2 - \left( 1 + \frac{1}{m^2} \right) \left( \left( \frac{b}{m} \right)^2 - r^2 \right)}}{1 + \frac{1}{m^2}}
\end{equation}

\begin{equation}
x_e = \frac{z_e - b}{m}
\end{equation}   

where:
\begin{align}
m &= \frac{z - z_i}{x - x_i} \quad \text{and} \quad
b = z - \left( \frac{z - z_i}{x - x_i} \right) \cdot x
\end{align}

Based on these intersection coordinates, the angle of intersection $\theta$ for a convex probe is calculated as follows:
\begin{equation}
\theta = \arctan \left( \frac{x_e}{z_e} \right)
\end{equation}

For a linear probe, the $x$ coordinate of the intersection point can be expressed as:
\begin{equation}
    {x_e}= \left(-\frac{{b}}{m}\right)
\end{equation}
 The resulting intersection points are then used to map the transmit aperture onto the compound mask. Since in the general case the intersection point does not correspond to the center of a probe element, an interpolation across adjacent probe elements on apodization weights is carried out to compute the corresponding compound weight.

\section{Results AND Discussion} 
\begin{figure*}[ht!]
    \centering

    \begin{minipage}{0.22\linewidth}
        \centering
        \textbf{(SALA)}
        \includegraphics[width=\linewidth]{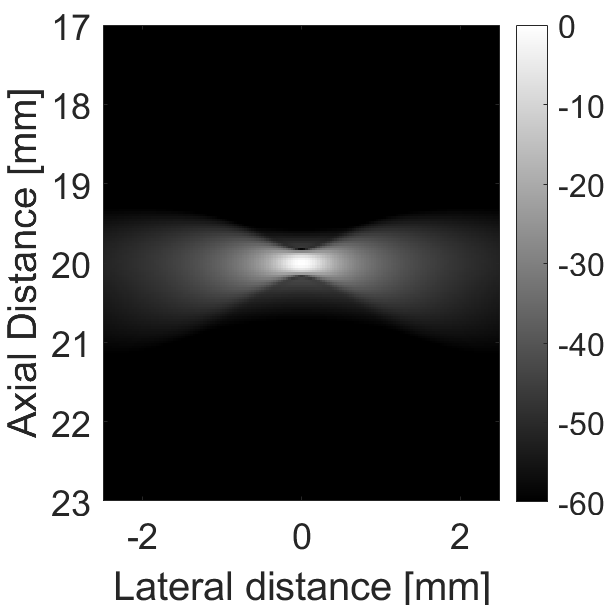}
    \end{minipage}
    \hfill
    \begin{minipage}{0.22\linewidth}
        \centering
        \textbf{(DWLA)}
        \includegraphics[width=\linewidth]{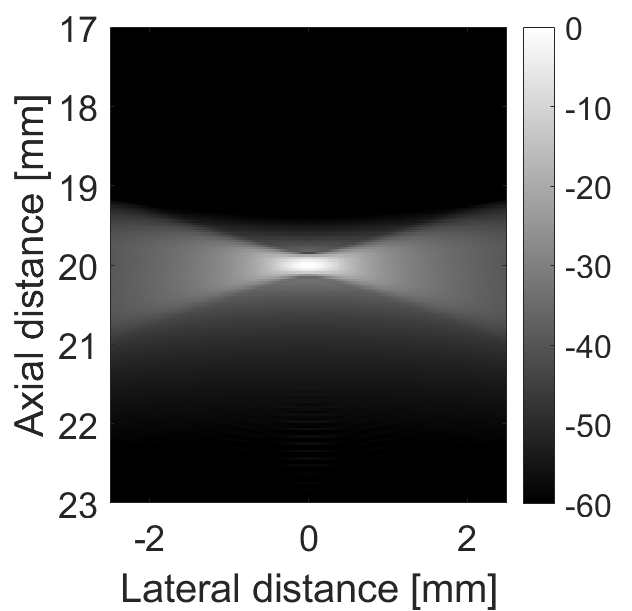}
    \end{minipage}
    \hfill
    \begin{minipage}{0.22\linewidth}
        \centering
        \textbf{(SACA)}
        \includegraphics[width=\linewidth]{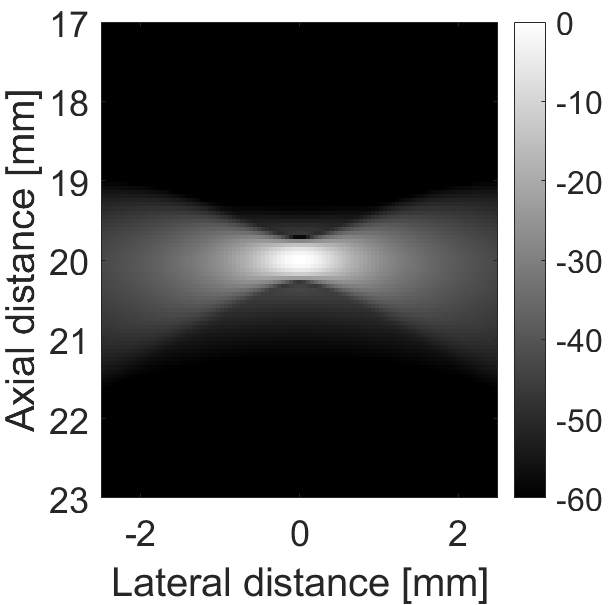}
    \end{minipage}
    \hfill
    \begin{minipage}{0.22\linewidth}
        \centering
        \textbf{(DWCA)}
        \includegraphics[width=\linewidth]{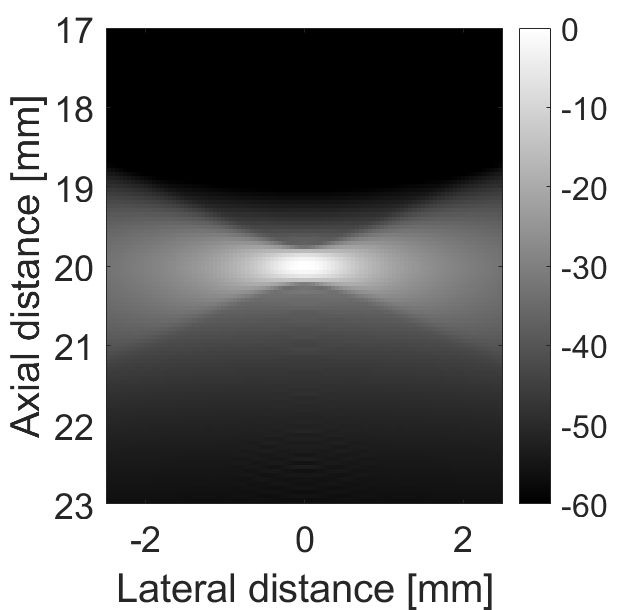}
    \end{minipage}

    \vspace{0.5cm} 

    \begin{minipage}{0.22\linewidth}
        \centering
        \includegraphics[width=\linewidth]{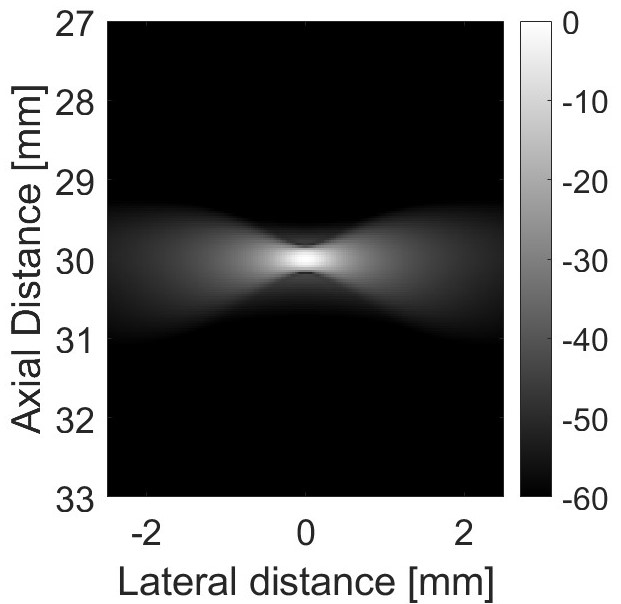}
    \end{minipage}
    \hfill
    \begin{minipage}{0.22\linewidth}
        \centering
        \includegraphics[width=\linewidth]{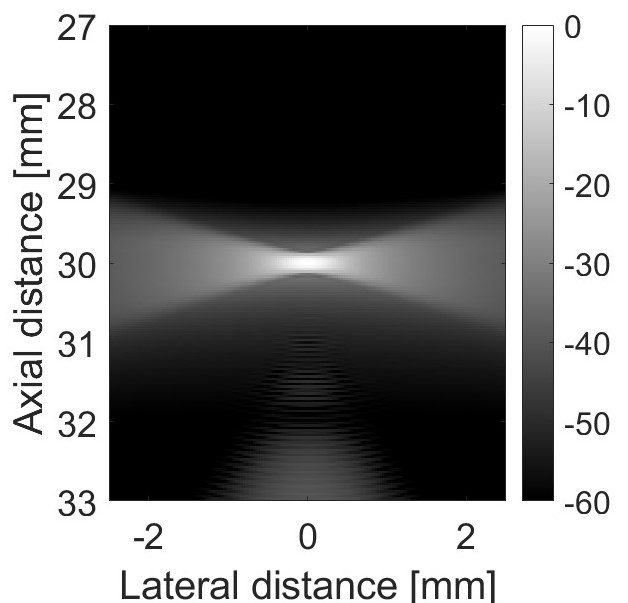}
    \end{minipage}
    \hfill
    \begin{minipage}{0.22\linewidth}
        \centering
        \includegraphics[width=\linewidth]{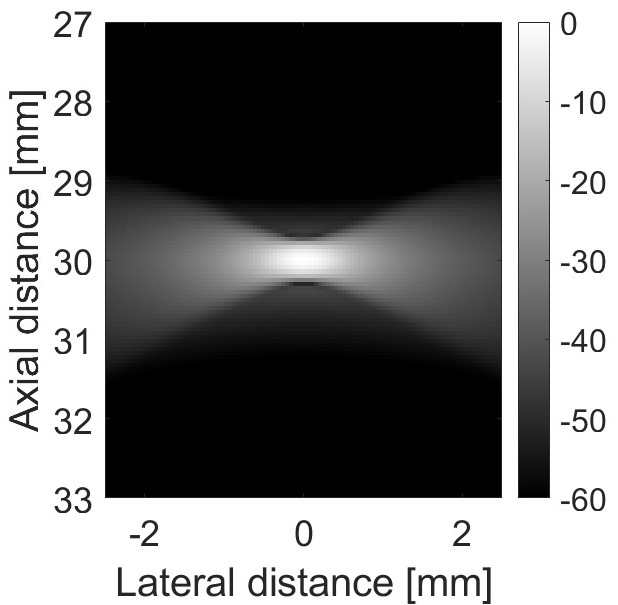}
    \end{minipage}
    \hfill
    \begin{minipage}{0.22\linewidth}
        \centering
        \includegraphics[width=\linewidth]{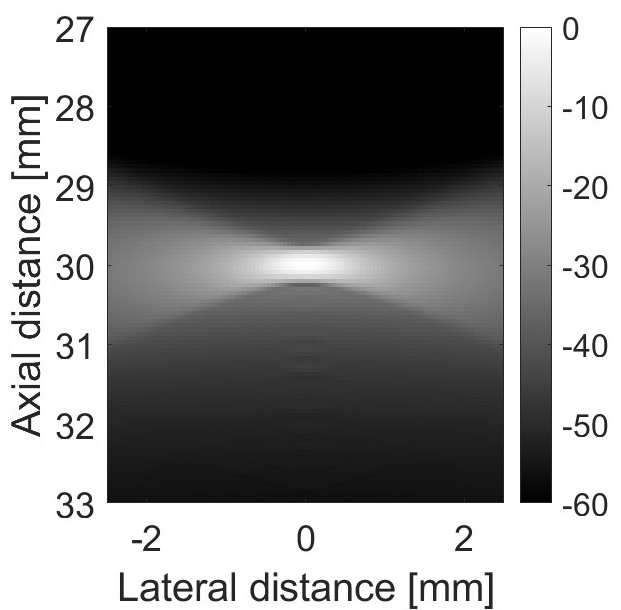}
    \end{minipage}

    \caption{PSF of SALA, DWLA, SACA, and DWCA  for a point at 20 mm depth (first row), and 30mm depth (second row) with F-number=0.7}
    \label{fig:psf-comparison}
\end{figure*}

\begin{figure*}[ht!]
    \centering

    \begin{minipage}{0.22\linewidth}
        \centering
        \textbf{(SALA)}
        \includegraphics[width=\linewidth]{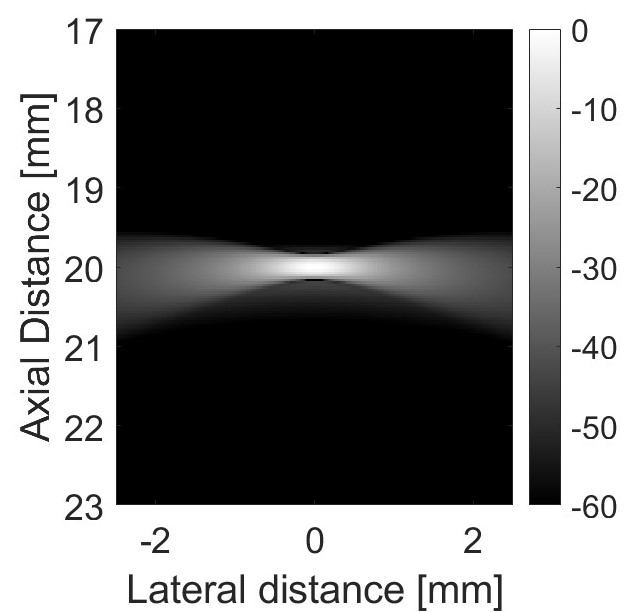}
    \end{minipage}
    \hfill
    \begin{minipage}{0.22\linewidth}
        \centering
        \textbf{(DWLA)}
        \includegraphics[width=\linewidth]{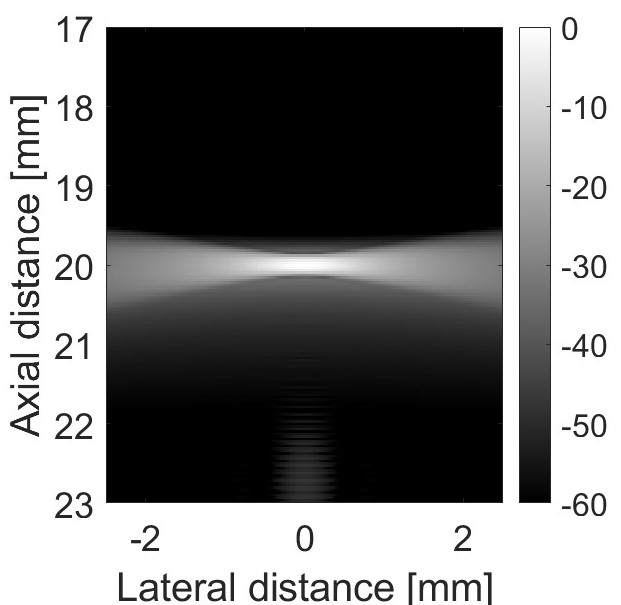}
    \end{minipage}
    \hfill
    \begin{minipage}{0.22\linewidth}
        \centering
        \textbf{(SACA)}
        \includegraphics[width=\linewidth]{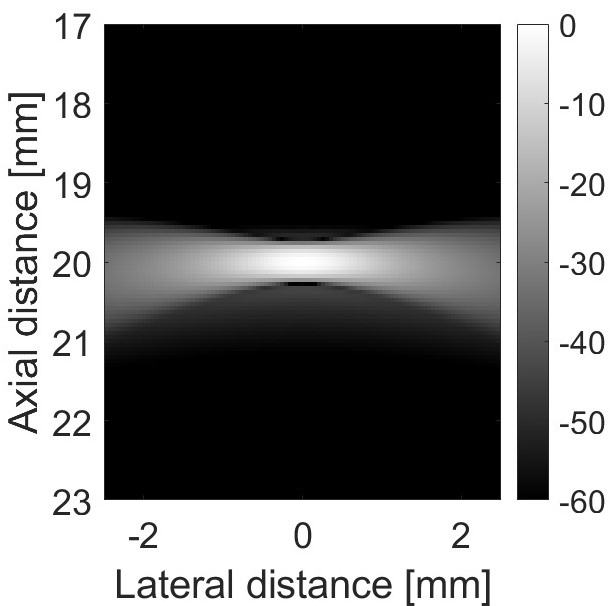}
    \end{minipage}
    \hfill
    \begin{minipage}{0.22\linewidth}
        \centering
        \textbf{(DWCA)}
        \includegraphics[width=\linewidth]{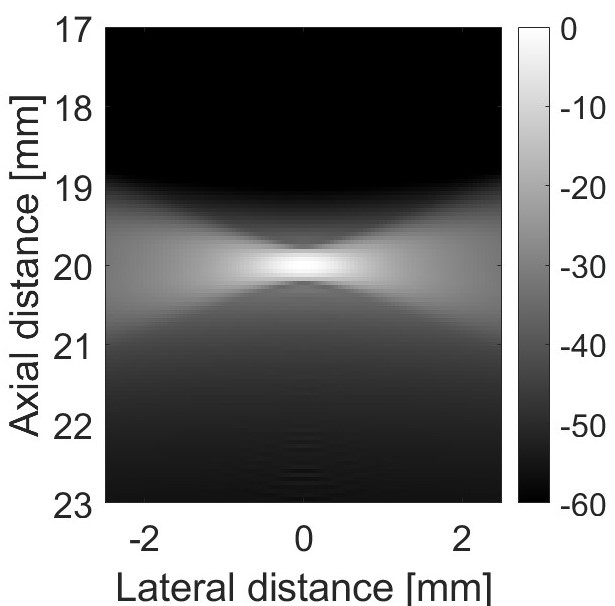}
    \end{minipage}

    \vspace{0.5cm} 

    \begin{minipage}{0.22\linewidth}
        \centering
        \includegraphics[width=\linewidth]{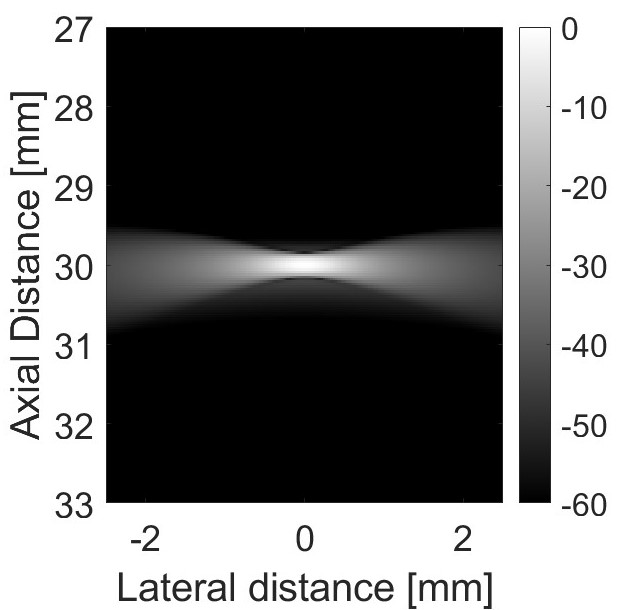}
    \end{minipage}
    \hfill
    \begin{minipage}{0.22\linewidth}
        \centering
        \includegraphics[width=\linewidth]{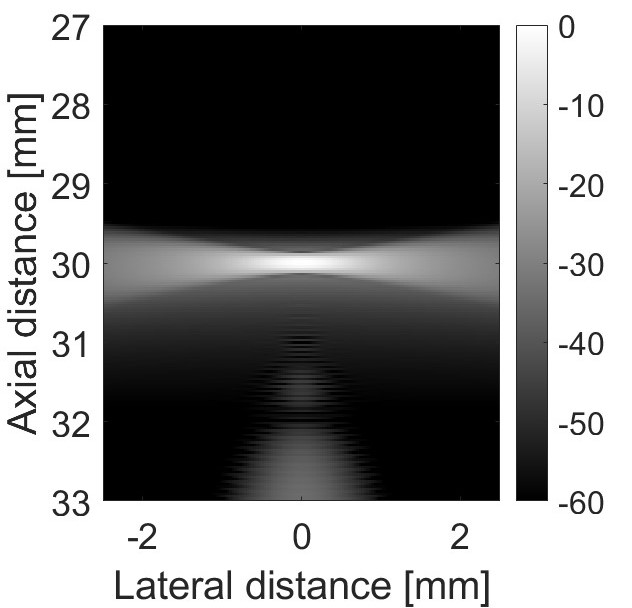}
    \end{minipage}
    \hfill
    \begin{minipage}{0.22\linewidth}
        \centering
        \includegraphics[width=\linewidth]{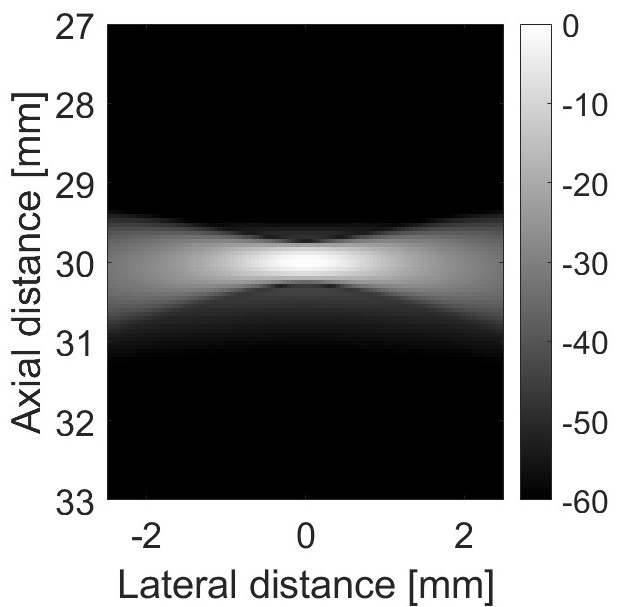}
    \end{minipage}
    \hfill
    \begin{minipage}{0.22\linewidth}
        \centering
        \includegraphics[width=\linewidth]{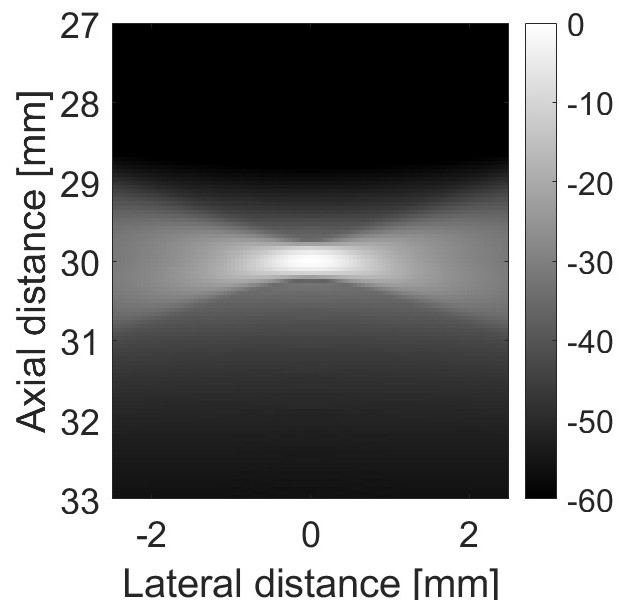}
    \end{minipage}

    \caption{PSF of SALA, DWLA, SACA, and DWCA  for a point at 20 mm depth (first row), and 30mm depth (second row) with F-number 1.4.}

\end{figure*}

The solution of the equations was validated through simulated data generated from the Field II software [12,13]  for both linear and convex probes. A 192-element array with a 5 MHz central frequency and 0.25 mm pitch was used for the linear probe, while a 3 MHz central frequency and 0.3 mm pitch were used for the convex probe with a radius of 50mm. A Hanning window was used in transmission and reception, with an F-number of 0.7 and 1.4 with a 2-cycle sinusoidal excitation signal. The image point spread function (PSF) for a point scatterer located at 20 mm and 30 mm depths was evaluated for the four cases, i.e. Synthetic Aperture Linear Array (SALA), Synthetic Aperture Convex Array (SACA), Divergent Wave Linear Array (DWLA), Divergent Wave Convex Array (DWCA).   For DWLA a linear distribution of virtual sources placed at a fixed distance behind the probe surface has been adopted. This arrangement is depicted in Figure 2(a), where the virtual sources are evenly spaced along the x-axis directly behind the probe elements. For DWCA the virtual sources were positioned along a circle behind the probe surface (Figure 2(b)). Research on optimized virtual source distributions for linear probes and convex arrays in 2D imaging is limited. Therefore, a deterministic distribution of virtual sources is employed for both probe types according to the proposal detailed in  [14]. For the linear probe, 20 virtual sources are uniformly distributed along the lateral axis, at a distance of 50 mm from the surface of the probe. These sources are positioned at regular intervals across a range of -25 to 25 times pitch. For the convex probe, the distribution of 20 virtual sources follows a curvilinear pattern to match the geometry of the probe elements. The angular range for these sources is defined by the maximum aperture angle set to $33^\circ$, covering a symmetric span on both sides of the probe's central axis. The curvilinear pattern of virtual sources is placed behind the probe, intersecting the probe center. 
To have an idea of how the transmit apodization weights are transformed into the compound mask, in Figure 4 the apodization weights function of probe position (left) and the compound mask function of virtual source index are displayed for a set of image points depths (i.e. the y coordinate of the displayed images),  for the linear probe case. As it can be seen the transformation consists in a geometric warping that locally preserves the numerical values. Figure 5 also illustrates the relationship between the transmit apodization used in synthetic aperture imaging (STAI) and the corresponding compound mask applied for divergent wave imaging (DWI), for a fixed image point. The left plot represents the transmit apodization applied to the lateral positions of the transducer elements during STAI. This apodization function demonstrates a smooth distribution of weights, peaking at the center and tapering off towards the edges, ensuring optimal energy distribution during transmission. The right plot depicts the derived compound mask for DWI, showing the corresponding weights as a function lateral position of the virtual sources (the virtual source depth is assumed to be fixed in this example). The derived mask closely mirrors the behavior of the STAI apodization but is adjusted for the spatial distribution of the virtual sources.

The images of point-like scatterers are displayed in Figure 6 and Figure 7 using F-number 0.7 and F-number of 1.4 respectively and their evaluation in terms of full width at half maximum (FWHM) is reported in Table I and table II. For linear probes, the Full Width at Half-Maximum (FWHM) with DWLA increased by 7.5\% at 20 mm and by 9\% at 30 mm compared to SALA. In contrast, DWCA showed a much smaller increase of 1.64\% at 20 mm and a substantial increase of 26.56\% at 30 mm compared to SACA, with a slight increase in sidelobes. Overall the proposed approach for the computation of the compound mask seems to reproduce quite accurately the effect of transmit apodization in Synthetic Aperture in terms of spatial resolution, except for the case of DWCA at 30 mm. 
This result accounts for the fact that the spatial distribution of virtual sources also plays a crucial role in image optimization, and it may limit the accuracy of mapping between STAI apodization and DWCA compounding. These aspects will be investigated in future studies. Concerning the side-lobe level, the worsening is likely due to the non-ideal nature of the divergent wavefield, given by the boundary effect introduced by the finite aperture.
\begin{table}[ht]
\centering
\begin{tabularx}{\columnwidth}{l *{2}{>{\centering\arraybackslash}X}}
\toprule
Technique & FWHM at 20 mm  & FWHM at 30 mm \\
\midrule
SALA & 0.40 & 0.43 \\
DWLA & 0.43 & 0.47 \\
SACA & 0.61 & 0.64 \\
DWCA & 0.62 & 0.81 \\
\bottomrule
\end{tabularx}
\caption{Comparison of FWHM values at 20 mm and 30 mm for different techniques  with F-number 0.7}
\end{table}
\begin{table}[ht]
\centering
\begin{tabularx}{\columnwidth}{l *{2}{>{\centering\arraybackslash}X}}
\toprule
Technique & FWHM at 20 mm  & FWHM at 30 mm \\
\midrule
SALA & 0.65 & 0.68 \\
DWLA & 0.8 & 0.95 \\
SACA & 0.8 & 0.84 \\
DWCA & 0.88 & 1 \\
\bottomrule
\end{tabularx}
\caption{Comparison of FWHM values at 20 mm and 30 mm for different techniques  with F-number 1.4}
\end{table}
\section{limitations and potential improvements}
This study primarily relies on simulations conducted using Field II software, but simulations alone cannot fully account for the complexities encountered in real-world scenarios, such as tissue heterogeneity. Therefore, experimental validation through in vitro or in vivo testing is necessary to comprehensively evaluate the effectiveness of our proposed method. Nevertheless, Field II has been extensively validated and is widely used within the ultrasound imaging community due to its accuracy in simulating wave propagation and evaluating imaging techniques. Moreover, numerous studies have shown that simulation-based results often correlate with in vivo findings for similar imaging methods [15, 16].
Another limitation of this study is that the proposed compound mask approach is designed to operate with transmit signals that are short in time domain. While this constraint simplifies the transformation and ensures accurate mapping of synthetic aperture imaging (SAI) transmit apodization weights to divergent wave imaging (DWI), it may limit the generalizability of the method to imaging setups or applications where longer transmit pulses are required. Also, the current approach does not explicitly account for the contribution of nearby elements when calculating the compound mask weights as a function of the apodization of a single probe element. Neighboring elements may have a significant impact on the transmitted and received signals.

One of the notable strengths of this study is the computation of optimized weights without the need for complex optimization algorithms, achieved by leveraging a closed-form approach. This method offers a streamlined solution that eliminates the necessity of iterative optimization processes, reducing computational overhead. Furthermore, the proposed approach introduces no additional computational load to divergent wave imaging (DWI). The compound mask weights can be efficiently computed offline and stored in memory, as demonstrated in [11], thereby minimizing the need for real-time computations and ensuring a resource-efficient implementation. Additionally, this study extends the application of the compound mask method beyond linear geometries by introducing its use for convex array configurations. This expansion represents a significant step forward in making the method applicable to a broader range of imaging scenarios, offering versatility and adaptability in high-frame-rate ultrasound imaging. Moreover, the presented mapping approach has the potential to be utilized in future research to optimize other imaging parameters, such as the distribution of virtual sources. By identifying optimal virtual source arrangements, it is possible to further enhance image quality and performance across various divergent wave imaging setups.

\section{Conclusion}

This study has demonstrated an approach to applying a compound mask in divergent wave imaging (DWI) by deriving a transformation from synthetic aperture imaging (SAI). The derived transformation enables using the same apodization window in synthetic transmit aperture imaging (STAI) for divergent waves (DW), effectively linking the compound mask with transmit apodization for each virtual source position. The results from simulated data using linear and convex probes validate the theoretical derivations. By examining the point spread function (PSF) at two different depths and comparing the full width at half-maximum (FWHM), it was found that the transformed apodization accurately replicates the effect of the STAI transmit apodization The results from simulated data confirm that the proposed compound mask method preserves the effect of transmit apodization in Synthetic Aperture Imaging (SAI) for divergent waves, as evidenced by a 7.5\% and 9\% increase in Full Width at Half Maximum (FWHM) at depths of 20 mm and 30 mm, respectively, for the Divergent Wave Linear Array (DWLA) compared to the Synthetic Aperture Linear Array (SALA). Similarly, for the Divergent Wave Convex Array (DWCA), a smaller increase of 1.64\% at 20 mm and a more pronounced increase of 26.56\% at 30 mm compared to the Synthetic Aperture Convex Array (SACA) was observed. These results highlight the ability of the proposed method to accurately replicate transmit apodization effects while optimizing compound mask weights in both linear and convex geometries. 

 Future research will focus on several key areas to extend and enhance the proposed approach. First, experimental validation using clinical datasets will be performed to assess the practical applicability and robustness of the compound mask method in real-world scenarios. Further optimization of divergent wave imaging parameters, such as virtual source positions, will be explored. By leveraging the mapping criteria developed in this study, we aim to identify parameters that can predict optimal virtual source distributions without generating point spread function (PSF) plots. This will enable a more intuitive and computationally efficient approach to position determination, potentially reducing trial-and-error in imaging setups. Another important area for future exploration is the adaptation and refinement of the compound mask technique for more complex imaging geometries, including 3D imaging configurations. Expanding the approach to three-dimensional probes could unlock new applications and improve image quality and resolution.

\begin{IEEEbiography}[]{Zahraa Alzein} was born in Lebanon. She received her Bachelor's degree in Electronics in 2018 and her Master's degree in Signal, Image Processing, and Telecommunications in 2020, both from the Lebanese University, Beirut, Lebanon. She is currently pursuing her Ph.D. in Electrical Engineering at the University of Genoa, Genoa, Italy, in collaboration with Esaote Company, focusing on advanced ultrasound imaging techniques.
She has worked on various projects during her studies, including the optimization of ultrasound beamforming techniques.  Her current research interests include focused transmit ultrasound systems and Ultrafast Imaging .
Ms. Alzein has been actively involved in IEEE conferences. She has received recognition for her work on computationally efficient ultrasound beamforming and continues to contribute to the advancement of medical ultrasound imaging technology.

\end{IEEEbiography}

\begin{IEEEbiography}[]{Marco Crocco} received the Laurea degree in
electronic engineering and the Ph.D. degree in
electronic engineering, computer science and
telecommunications from the University of Genova,
Genova, Italy, in 2005 and 2009, respectively.
From 2005 to 2010, he was with the Department
of Biophysical and Electronic Engineering,
University of Genova. From 2010 to 2015, he
joined as a Post-Doctoral Fellow with the Pattern
Analysis and Computer Vision Department and the
Visual Geometry and Modeling Laboratory, Istituto
Italiano di Tecnologia, Genova, Italy. He is currently working on signal processing applied to medical ultrasound in Esaote S.p.A.
\end{IEEEbiography}

\begin{IEEEbiography}[]{DANIELE D. CAVIGLIA}(Life Member, IEEE)
is currently a Full Professor in electronics
with the Department of Electrical, Electronics
and Telecommunications Engineering and Naval
Architecture (DITEN), University of Genoa, Italy,
where he teaches courses of electronic systems
for telecommunications with the Laboratory of
Electronics and Embedded Systems. His current
research interests include the design of electronic
circuits and systems for telecommunications, electronic equipment for health and safety, and energy harvesting techniques for
Internet of Things (IoT) applications. He is also active in the development
of innovative solutions for environmental monitoring and this activity led to
the foundation of Artys srl as a spin-off of the University of Genoa, in 2014.

\end{IEEEbiography}

\EOD

\end{document}